\documentclass[pdflatex,sn-basic]{sn-jnl}

\usepackage{graphicx}%
\usepackage{multirow}%
\usepackage{amsmath,amssymb,amsfonts}%
\usepackage{amsthm}%
\usepackage{mathrsfs}%
\usepackage[title]{appendix}%
\usepackage{xcolor}%
\usepackage{textcomp}%
\usepackage{manyfoot}%
\usepackage{booktabs}%
\usepackage{algorithm}%
\usepackage{algorithmicx}%
\usepackage{algpseudocode}%
\usepackage{listings}%
\usepackage{revsymb}

\begin{document}

\title[Fractional Stars]{Fractional Stars}


\author[1]{\fnm{Hooman} \sur{Moradpour}}\email{h.moradpour@riaam.ac.ir}

\author[2,3]{\fnm{Shahram} \sur{Jalalzadeh}}\email{shahram.jalalzadeh@ufpe.br}

\author[1]{\fnm{Mohsen} \sur{Javaherian}}\email{javaherian@maragheh.ac.ir}

\affil[1]{\orgdiv{Research Institute for Astronomy and Astrophysics of Maragha (RIAAM)}, \orgname{University of Maragheh}, \orgaddress{
\city{Maragheh}, \postcode{55136-553},
\country{Iran}}}

\affil[2]{\orgdiv{Departamento de F\'{i}sica}, \orgname{Universidade Federal de Pernambuco}, \orgaddress{
\city{Recife-PE}, \postcode{50670-901},
\country{Brazil}}}

\affil[3]{\orgdiv{Department of Physics and Technical Sciences}, \orgname{Western Caspian University}, \orgaddress{
\street{AZ 1001},
\city{Baku}, 
\country{Azerbaijan}}}


\abstract{This study examines the possibility of starting the process of
collapsing and forming stars from a fractional molecular cloud.
Although the Verlinde's approach is employed to derive the
corresponding gravitational potential, the results are easily
generalizable to other gravitational potential proposals for
fractional systems. It is due to the fact that the different
methods, despite the difference in the details of results, all
obtain power forms for the potential in terms of radius. An
essential result of this analysis is the derivation of the
corresponding Jeans mass limit, which is a crucial parameter in
understanding the formation of stars. The study shows that the
Jeans mass of a cloud in fractional gravity is much smaller than
the traditional value. In addition, the study also determines the
burning temperature of the resulting star using the Gamow theory.
This calculation provides insight into the complex processes that
govern the evolution of these celestial bodies. Finally, the study
briefly discusses the investigation of hydrostatic equilibrium, a
crucial condition that ensures the stability of these fractional
stars. It also addresses the corresponding Lane--Emden equation,
which is pivotal in understanding this equilibrium.}




\maketitle

\section{Introduction}\label{sec1}

The primary objective of fractional calculus is to broaden the
scope of classical (ordinary) calculus by incorporating the
principles and techniques of differentiation and integration to
encompass non-integer orders. In classical calculus,
differentiation and integration are limited to integer orders like
first, second, or third and do not consider non-integer or
irrational orders. However, the emergence of fractional calculus
has transformed this conventional approach, allowing for the
exploration of orders beyond integer values and even into the
realm of irrational numbers \citep{Herrmann2014,
Jalalzadeh:2020bqu,pozrikidis2018fractional}. One of the primary
incentives behind the study of fractional calculus lies in the
necessity to effectively represent and examine phenomena that
manifest fractal characteristics or possess fractal structures
\citep{fractaldimension}. Fractal geometry serves to elucidate
objects or processes that demonstrate self-similarity across
varying magnitudes (interested readers can consult
\citep{Achwanden2011self, fractalastro, Roy:2009tr,
Majumder:2017rsy, Tajik2023} for some astronomical applications).
Fractional calculus, in this regard, offers a mathematical tool
that enables the description and analysis of such phenomena.
Furthermore, it is worth noting that fractional calculus
establishes connections with probability theory, signal and image
processing, and control theory \citep{Tarasov2022,
Duarte2020signal, Chen2019image}. Using fractional derivatives and
integrals facilitates the modelling and analysis of random and
non-stationary signals while also contributing to developing more
advanced controllers for intricate systems \citep{13}.

It is worth noting that the study of fractional calculus is driven
by the need for more precise models that can accurately explain
the behaviour of various physical and engineering problems.
Several processes in nature and engineering exist that
integer-order derivatives and integrals cannot effectively
describe. For instance, in the case of viscoelastic materials, the
stress-strain relationship may require fractional derivatives to
represent the behaviour \citep{Zheng2017} accurately. In fact,
fractional calculus aims to develop a comprehensive mathematical
framework that allows for studying and analyzing systems and
phenomena that go beyond classical calculus. By extending the idea
of differentiation and integration to non-integer orders,
fractional calculus provides a valuable tool for modelling and
understanding complex behaviours in various scientific and
engineering disciplines \citep{Herrmann2014}.

The integration of fractional differential equations has proven to
be a precious tool in accurately describing and solving complex
real-life problems that are encountered in the fields of particle
physics, wave mechanics, electrical systems, fractal wave
propagation, and numerous other areas of study
\citep{Stanislavsky2010, Herrmann2014}. Moreover, extensive
research has shown that using fractional derivatives can
significantly enhance the precision and effectiveness of the
detection criterion in signal processing \citep{Mathieu2003}. This
development has opened up new avenues for advancements in the
field of signal processing, providing researchers with a powerful
tool to tackle challenging problems in various domains.

There are several applications of fractional calculus in gravity
and cosmology, which are currently being actively researched. The
use of fractional calculus has proved effective in addressing
gravitational forces and cosmological models, making it an
essential tool for scientific inquiry \citep{Costa:2023zfg,
Garcia-Aspeitia:2022uxz, Shchigolev:2021lbm, Jalalzadeh:2022uhl,
Leon:2023iaq, Gonzalez:2023who, Socorro:2023ztq, Debnath2012,
Momeni2012, Socorro:2023xmx}. These applications include the
stochastic gravitational wave background in quantum gravity
\citep{Calcagni:2020tvw}, gravitational-wave luminosity distance
\citep{Calcagni:2019ngc}, inflation and CMB spectrum
\citep{Rasouli:2022bug, Calcagni:2017via}, fractional action
cosmology \citep{El-Nabulsi:2012wpc, Jamil:2011uj},  discrete
gravity \citep{El-Nabulsi:2013mma}, non-minimal coupling and
chaotic inflation  \citep{El-Nabulsi:2013mwa}, phantom cosmology
with conformal coupling \citep{Rami:2015kha}, Ornstein--Uhlenbeck
like fractional differential equation in cosmology
\citep{El-Nabulsi:2016dsj}, fractional action cosmology with a
variable order parameter \citep{El-Nabulsi:2017vmp}, non-commutative classical and quantum fractionary Cosmology of FRW Case \cite{Socorro:2024poa},and wormholes
in fractional action cosmology \citep{El-Nabulsi:2017jss}.
Additionally, new metrics and dark energy models in emergent,
logamediate, and intermediate models of the Universe have been
considered \citep{Debnath2012}. \citet{Shchigolev:2010vh,
Shchigolev:2013jq} have obtained several exact solutions for
cosmological models, which significantly deviate from the standard
model \citep{Calcagni:2009kc} due to the fractal nature of
space-time. The interval $1\leq\alpha <2$ (the Lévy’s
fractional parameter) is explored in Ref.
\citep{Jalalzadeh:2022uhl} using Riesz's fractional derivative to
acquire non-boundary and tunnelling wave functions for a closed de
Sitter geometry. The pre-inflation epoch is investigated in Ref.
\citep{Rasouli:2022bug} within the framework of fractional quantum
cosmology. The thermodynamics of fractional black holes (BH) is
studied in Ref. \citep{Jalalzadeh:2021gtq}. Similarly, the
Friedmann and Raychaudhuri equations are modified using fractional
calculus to examine the dynamics of the Universe in the absence of
cold dark matter (CDM) and dark energy \citep{Barrientos:2020kfp}. Friedmann equations of the fractal apparent horizon are analyzed in \cite{Jalalzadeh:2024qej}.
Additionally, fractional calculus is employed to determine the
value of the cosmological constant, which needs to be restructured
due to the well-known ultraviolet divergence in traditional
quantum field theory \citep{Landim:2021www, Landim:2021ial}. The
fractional approach is also explored in the context of modified
Newtonian dynamics (MOND) and quantum cosmology in
\citet{Giusti:2020rul,Torres:2020xkw,Barrientos:2020kfp}.

The Lane-Emden equation has a rich history in mathematical physics
and astrophysics and plays a crucial role in understanding various
phenomena. These phenomena include but are not limited to the
theory of stellar structure, the behaviour of spherical gas clouds
in terms of their thermal properties, the study of isothermal gas
spheres, and the investigation of thermionic currents
\citep{Gamowbook}. In an effort to expand upon the Lane--Emden
equation further, researchers have explored the application of
fractional calculus. By doing so, they have successfully
introduced a fractional Lane--Emden equation, which in turn
generates a family of Emden--Fowler differential equations
\citep{ElNabulsi2017}. The interested reader can delve deeper into
applying this particular type of equation in a polytropic gas
sphere by referring to the comprehensive work of
\citep{ABDELSALAM2020} and the references provided therein.
Additionally, researchers have made numerous attempts to find
fractional Newtonian potentials and explore their potential in
modelling the MOND theory \citep{Varieschi:2020hvp,
Varieschi:2020ioh, FractalLaplac, Varieschi:2022mid,
Calcagni:2021mmj, Benetti:2023nrp, virialfractal} by introducing a
fractional Poisson equation or utilizing Gauss law. These
endeavours highlight the ongoing efforts to expand our
understanding of the Lane--Emden equation and its applications in
various areas of physics and astrophysics. A molecular cloud
begins to collapse if it has enough gravitational energy to
overcome the kinetic energy of its constituents. The threshold
mass for having a collapse is called the Jeans mass which is an
important criterion to study the astrophysical applications of
gravitational theories \citep{Forgan:2011uc, Capozziello:2011gm,
Vainio:2015ejx, Bessiri:2021kug, Lima:2001vq, Moradpour:2019wpj,
Ourabah:2020qzy}. Of course, it seems that Bok globules do not
follow this law \citep{Vainio:2015ejx}. Thus, it is a challenge to
find a theory reducing the value of ordinary Jeans mass (deduced
from Newtonian gravity and standard extensive statistical
mechanics and thermodynamics) so that the star formation in Bok
globules is justified \citep{Moradpour:2019wpj}. On the other
hand, the temperature of stars was a puzzle solved by George Gamow
\citep{Gamowbook}. Just the same as the Jeans mass, the sample
statistics have a vital role in deriving the Gamow temperature
\citep{MoradpourJava, Ourabah:2023byw}.

Our primary objective is to present an initial explanation
regarding the star formation process within a molecular cloud by
utilizing fractional gravity. The resulting equations serve as
essential prerequisites for star formation and further serve to
validate and support the concept of star equilibrium. The
subsequent section will examine the fractional Jeans mass within a
fractional molecular cloud to accomplish this objective. In this
study, we will cover the following topics. In section 2, we will
obtain the fractional gravitational potential utilizing the
Verlinde's entropic force. In the third section, we will discuss
the properties of fractional molecular gas and their role in
forming fractional stars. In Section Four, we will focus on
determining the burning temperature of a fractional star, which
results from the contraction of a fractional molecular gas. In
Section 5, we will derive the Lane--Emden equation that
corresponds to this system. Finally, the last section will
summarize the key points and findings discussed throughout the
study.

\section{Fractional gravitational potential}

To incorporate fractional mechanics into the physics of cloud
collapsing, it is necessary to expand upon the fractional
gravitational potential, denoted as $V(r)$. Consequently, we will
utilize Verlinde's entropic force conjecture\footnote{Entropic
gravity, also known as emergent gravity, is a concept that
proposes a depiction of gravity as a force arising from entropy.
While this force displays uniformity on a larger scale, it is
susceptible to disorder at the quantum level, setting it apart
from being a fundamental interaction \citep{Verlinde:2010hp}.}  to
determine the gravitational potential. To obtain the Newtonian
potential, we will follow Verlinde's entropic force conjecture
\citep{Verlinde:2010hp}, which has been utilized by
various authors (see for example \citep{Moradpour:2019wpj,
Moradpour:2017fmq, Sheykhi:2019bsh, shey, Moradpour:2024azo}) and
references therein to explore modifications of Newtonian
dynamics, MOND theory, and Jeans mass by considering different
entropies.

In this section, we will explore the fractional Newtonian
potential. To achieve this, we will first review the original
Verlinde's entropic force conjecture to obtain ordinary Newtonian
gravitational potential in subsection \ref{sub1}, which will then
be extended to the fractional case in subsection \ref{sub3}.
Additionally, we will extend the entropy of fractional BH to an
arbitrary holographic screen in subsection \ref{sub2}.

\subsection{Derivation of the gravitational potential utilizing Verlinde's entropic force}\label{sub1}

Verlinde proposed the entropic force to explain Newton's second
law and Einstein's field equations. This revolutionary idea
suggests that gravity can be understood as an emergent force that
arises from the statistical behaviour of microscopic particles
rather than being an inherent force. The concept of an entropic
force has been applied not only to the distribution of quantum
particles but also to the derivation of quantum osmotic pressure
and the resulting entropic forces for both bosonic and fermionic
particles. The core concept of this approach is founded upon two
fundamental principles. The first principle, the holographic
principle, postulates that space itself is not a standalone entity
but a medium that contains and encapsulates information. This
information is intricately stored within holographic screens,
which act as boundary surfaces separating various points. These
screens are constructed using potentials derived from time-like
Killing vectors, resulting in equipotential surfaces. Moving on to
the second principle, it proposes that when an object or body
undergoes motion relative to the holographic screen, it exerts a
force that can be likened to the entropic force observed in
phenomena such as osmosis or the interaction between large colloid
molecules and smaller particles in a thermal environment.
Consequently, according to the Verlinde approach,  gravity is not
categorized as a fundamental interaction akin to the other three
known forces but manifests as a type of entropic force, so that
force is a net result of the tendency of systems to increase their
entropy. \citet{Verlinde:2010hp} proposed the entropic force as an
explanation for gravity. This idea suggests that gravity arises
from the statistical behaviour of microscopic particles. The
approach is founded upon the holographic principle and the
principle of motion. According to this approach, gravity is not
categorized as a fundamental interaction but manifests as a type
of entropic force.

Verlinde's theory states that when a test particle of mass $m$
moves away from a holographic screen, it experiences an entropic
force given by the equation

\begin{equation}
    \label{SS1}
    F=-T\frac{\Delta S}{\Delta x},
\end{equation}
where $\Delta x$ is the particle's displacement from the screen,
$T$ is the temperature, and $\Delta S$ is the entropy change on
the screen. In this theory, the entropy of a BH plays
a crucial role in deriving Newton's law of gravity through the
entropy-area relationship, where
$S_\text{BH}=k_\text{B}A/4L_\text{P}^2$. Verlinde's approach
assumes that a holographic screen encloses any energy source, and
since the horizon of a BH is a holographic screen, the entropy of
a holographic screen is equal to the entropy of a BH. Thus, the BH
entropy can be used to derive gravitational force as the final
result of the tendency of systems to increase their entropy, and
spacetime and gravity are emergent phenomena. Indeed, this
assumption is also strengthened by Ref.~\cite{Srednicki:1993im}
where it is shown that if the ordinary Von-Neumann entropy is
employed, then the entropy of the boundary surface (holographic
screen) between two particles is equal to the Bekenstein entropy
(the entropy of corresponding BH).

We have two masses: a test mass ($m$) and a source mass ($M$) at the centre of a spherical surface $\Sigma$. When the test mass is placed close to the surface, it modifies the surface's entropy by one fundamental unit ($\Delta S$). The surface area's discrete spectrum determines this modification and is proportional to the test mass's Compton wavelength. Using the Bekenstein--Hawking entropy formula, which calculates BH entropy ($S=S(A)$), we get $\Delta S=\frac{\partial S}{\partial A}\Delta A$.


The energy of a surface $\Sigma$ is equivalent to the rest mass
energy of the source mass ($E=Mc^2$). The surface area is
quantized in units of the Planck area, and there are a set of
``bytes'' of information on it proportional to its area
($A/L_\text{P}^2=\gamma N$). The temperature of the surface can be
calculated using $T=\frac{2Mc^2}{k_\text{B}N}=\frac{\gamma
MG\hbar}{2\pi k_\text{B}cR^2}$, where $k_\text{B}$ is the
Boltzmann constant, $G$ is the gravitational constant, $\hbar$ is
the reduced Planck constant, and $R$ is the radius of the surface.

Thus, the gravitational force is an entropic force given by
\begin{eqnarray}\label{6}
F=-T\frac{\Delta S}{\Delta x}=-T\frac{\Delta A}{\Delta x}\frac{dS}{dA}=-\frac{mMG\gamma^2}{2\pi\eta R^2}\cdot\frac{L_\text{P}^2}{k_\text{B}}\cdot \frac{dS}{dA}.
\end{eqnarray}
Using Bekenstein--Hawking entropy
\begin{equation}
    \label{SS4}
    S_\text{BH}=k_\text{B}\frac{A}{4L_\text{P}^2},
\end{equation}
and choosing $\gamma^2={8\pi}\eta$, we obtain the Newtonian gravitational force and, consequently, the corresponding gravitational potential. 

\subsection{From the spectrum of a fractional BH to the fractal horizon surface }\label{sub2}

In the preceding subsection, we observed the importance of
possessing a compact holographic screen and the entropy of a BH in
Verlinde's reasoning to obtain Newtonian gravitational potential.
Therefore, to extend Verlinde's proposal to the fractional
situation, our initial action requires a thorough investigation of
the fundamental constituents of the fractional entropy of the
Schwarzschild BH.

The fractional Wheeler--DeWitt equation gives the following BH's
mass spectrum \citep{Jalalzadeh:2021gtq}
\begin{equation}
   \label{SS7}
  M=\left(\frac{\Omega_{d-1}}{\Omega_d}n\pi\right)^\frac{1}{d}M_\text{P},~~~~n=\{\text{large positive integer}\},
\end{equation}
where $M_\text{P}$ is the Planck mass, $\Omega_d$ is the volume of a $d$-dimensional unit sphere, and $d$ is given by the Lévy’s fractional parameter $\alpha$ as\footnote{Lévy's fractional parameter is crucial in understanding stable distributions and Lévy processes. It falls within the range of 0 to 2, with $\alpha=2$ corresponding to a Gaussian distribution. Stable distributions are invariant under convolution operations. In fractional quantum mechanics, the parameter $\alpha$ articulates the fractal dimension of trajectories. Laskin extended the Feynman path integral by using Lévy paths, involving $1<\alpha\leq2$. In space fractional quantum mechanics, the existence of the expectation value of the position and momentum of
a particle demands that the moments of first order exist. In this regard, $\alpha$ has to be restricted to the range $1< \alpha \leq 2$ \citet{Laskin:2002zz}. In the realm of quantum gravity, it is widely accepted that the expectation values of metric components and their corresponding momenta do not have a clear definition \cite{Jalalzadeh:2020bqu}. Also, these quantities do not meet the requirements to be considered as Dirac observables, or in most cases the wave function is not square-integrable. Consequently, the constraint on the value of $\alpha$ is not applicable in this context.}
\begin{equation}
   \label{SS8}
    d=\frac{2}{\alpha}+1
\end{equation}
As depicted in \citet{Jalalzadeh:2021gtq}, the aforementioned fractional mass spectrum of the BH directs us toward the subsequent fractional entropy
\begin{equation}
    \label{SS9}
    S_\text{fractal-BH}=S_\text{BH}^\frac{d}{2},
\end{equation}
where $S_\text{BH}$ is the entropy of ordinary BH in Eq. (\ref{SS4}).

This result shows that one can rewrite the Eq. (\ref{SS9}) as
\begin{equation}
    \label{FB1}
    S_\text{fractal-BH}=\frac{A_\text{fractal}}{A_\text{P}}.
\end{equation}
where the fractal surface of the BH horizon is defined by \citet{Junior:2023fwb}
\begin{equation}
    \label{FB2}
    A_\text{fractal}=A_\text{P}\left(\frac{A_\text{S}}{A_\text{P}}\right)^\frac{d}{2}=4\pi^\frac{d}{2}L_\text{P}^{2-d}R_\text{S}^d,
\end{equation}
where $A_\text{S}=16\pi G^2M^2$ is the Schwarzschild area of the
BH, with corresponding radius $R_\text{S}=2MG$. In addition, as it
is shown in \citet{Junior:2023fwb}, the above equations lead us to
conclude that the horizon of fractional BH is a fractal surface
with a dimension equal to $d=2/\alpha+1$, as defined in
Eq. (\ref{SS8}). Regarding L\'{e}vy's fractional parameter, $\alpha$ is restricted to the interval $1 <\alpha \leq 2$ (see, for example, \citet{Laskin:2002zz}) in fractional quantum mechanics. Then, if we assume the existence of the same restriction on the value of $\alpha$ in quantum gravity, then the fractal dimension of fractal-fractional BH falls into the following interval
\begin{equation}
    \label{FB4}
    2\leq d<3.
\end{equation}

Employing the fractal BH entropy (\ref{FB1}), one can define effective Schwarzschild radius, $R_\text{eff}$ and mass, $M_\text{eff}$ as
\begin{equation}\label{FB5}
    \begin{split}
   R_\text{eff}&=(4\pi)^\frac{d-2}{4}\left(\frac{R_\text{S}}{L_\text{P}}\right)^\frac{d}{2}L_\text{P}.\\
   M_\text{eff}&=\frac{R_\text{eff}}{2G}=(4\pi)^\frac{d-2}{4}\left(\frac{M}{M_\text{P}}\right)^\frac{d}{2}M_\text{P}.
    \end{split}
\end{equation}
As a result, one can rewrite the fractal surface area and fractal entropy of the BH as
\begin{equation}
    \label{FB6}
    \begin{split}
        A_\text{fractal}&=4\pi R^2_\text{eff},\\
        S_\text{fractal-BH}&=\frac{A_\text{fractal}}{4G}=\frac{\pi}{G}R_\text{eff}^2.
    \end{split}
\end{equation}
Substituting the fractional mass spectrum (\ref{SS7}) into the above definition of the fractal surface gives the spectrum of the fractal horizon surface
\begin{equation}
    \label{SS12}
    A_\text{fractal}(n)=(16\pi)^\frac{d}{2}\frac{\pi\Omega_{d-1}}{\Omega_d}L_\text{P}^2n,~~~~n=\{\text{Large integer numbers}\}.
\end{equation}
Therefore, if we assume on the surface $\Sigma_\text{fractal}$, there lives a set of `bytes' of information that scale proportional to the area of the fractal surface so
that $A_\text{fractal}/L_\text{P}^2=\gamma N$, where $N$ represents the number of bytes of information on $\Sigma_\text{fractal}$, then
\begin{equation}
    \Delta A_\text{fractal}= A_\text{fractal}(n+1)- A_\text{fractal}(n)=\gamma L_\text{P}^2,~~~\Delta N=(n+1)-n=1,
\end{equation}
and we obtain
\begin{equation}
    \gamma=(16\pi)^\frac{d}{2}\frac{\pi\Omega_{d-1}}{\Omega_d}.
\end{equation}

\subsection{Verlinde's entropic force and gravitational potential in the presence of the fractal holographic screen}\label{sub3}

We have completed all the necessary preparations and can now
further explain Verlinde's initial proposal. Our approach includes
considering a fractal holographic screen, which will play an
essential role in our attempt to derive the Newtonian
gravitational potential.

Suppose we have two masses. One is a test mass, denoted as $m$,
while the other is a source mass, denoted as $M$. As discussed in
subsection \ref{sub1}, the source mass $M$ is positioned at the
centre, and around it is a fractal surface denoted as
$\Sigma_\text{Fractal}$. Let us assume that the test mass is near
the surface relative to its reduced Compton wavelength.

Regarding our results in the previous subsection, it can be
inferred that a spherical surface exhibiting fractional-fractal
properties, $\Sigma_\text{Fractal}$, possesses an effective radius
that is defined in a manner similar to relation (\ref{FB5})
\begin{equation}
    \label{SS10a}
    \tilde r=(4\pi)^\frac{d-2}{4}\left(\frac{r}{L}\right)^\frac{d}{2}L,
\end{equation}
where $r$ denotes, as usual, the radial coordinate and $L$ denotes
the grid size. The effective surface area of
$\Sigma_\text{Fractal}$, then will be $A_\text{fractal}=4\pi\tilde
r^2$. Repeating the same assumptions of the first part of the
section for a test particle with a source mass $M$ at the centre
of a fractal sphere $\Sigma_\text{Fractal}$, we find \footnote{We
assumed that in a fractional-fractal situation, the equipartition
law, $A_\text{fractal}/L^2=\gamma N$, and $\Delta
x=\eta\lambdabar$ are accurate.}
\begin{equation}\label{6fractal}
\begin{split}
F=&T_\text{fractal}\frac{\Delta S_\text{fractal}}{\Delta x}=T_\text{fractal}\frac{\Delta A_\text{fractal}}{\Delta x}\frac{dS_\text{fractal}}{dA_\text{fractal}}\\
=& -\frac{mMG\gamma^2}{2\pi\eta \tilde r^2}\cdot\frac{L^2}{k_\text{B}}\cdot \frac{dS_\text{fractal}}{dA_\text{fractal}},
\end{split}
\end{equation}
where
\begin{equation}
    \label{fractalT}
    T_\text{fractal}=\frac{2Mc^2}{k_\text{B}N_\text{fractal}}=\frac{\gamma MG\hbar}{2\pi k_\text{B}c\tilde r^2}.
\end{equation}
Choosing $\eta=\gamma^2/8\pi$ and using (\ref{FB1}) lead us to the fractional modification of the Newtonian gravitational force
\begin{equation}
    \label{SS13}
    F=-\frac{GmM}{\tilde r^2}=-\frac{G_dmM}{r^d},
\end{equation}
where
\begin{equation}
    \label{SS14}
    G_d=\frac{GL^{d-2}}{(4\pi)^\frac{d-2}{2}},
\end{equation}
is the effective gravitational constant modified with quantum
gravity effects. Now, using the fractional form of the force in
(\ref{SS13}), it is easy to verify that the fractional
gravitational potential is given by
\begin{equation}
    \label{SS15}
    V(r)=-\frac{G_dM}{d-1}\frac{1}{r^{d-1}}.
\end{equation}

It is worth noting that several studies have explored similar
potentials that differ in the power of $r$ and the coefficient
$\frac{G_dM}{d-1}$
\citep{Muslih2007,Giusti:2020kcv,Varieschi:2020ioh,Varieschi:2022mid,Benetti:2023nrp}.
The differences stem from the methods employed to include the
fractional considerations.


\section{Jeans mass}
Consider a low-density cloud that maintains a uniform temperature $T$ and is in a state of hydrostatic equilibrium. Our objective is to deduce the condition for stability in a specific volume, denoted by $V$, which can be assumed to be spherical for simplicity and contains a given mass $M$. Let us represent the radius (the characteristic length) of the volume $V$ as $R$ and the total number of particles inside the volume as $N$, where each particle has a mass of $m$. {For the occurrence of collapse to occur, it is essential that the absolute value of the gravitational potential energy surpasses that of the thermal kinetic energy possessed by the cloud. Jeans Mass, $M_\text{J}$, which represents the critical threshold at which gravitational collapse occurs due to self-gravity, can be found as \citet{Gamowbook}
\begin{eqnarray}\label{3}
M_\text{J}\equiv\left(\frac{5k_\text{B}T}{Gm}\right)^{\frac{3}{2}}\left(\frac{3}{4\pi\rho_0}\right)^{\frac{1}{2}},
\end{eqnarray}
where $\rho_0 = \ \frac{3M}{4\pi R^3}$ denotes the gas density. If a particular area within the cloud possesses a mass that exceeds the Jeans mass, it will experience instability and undergo gravitational collapse \citep{Gamowbook}.}
%
%
%

The alteration in the value of $M_\text{J}$ undoubtedly varies according to the gravitational theory and statistical mechanics considered for examination \citep{Forgan:2011uc, Capozziello:2011gm, Vainio:2015ejx, Bessiri:2021kug, Lima:2001vq, Moradpour:2019wpj, Ourabah:2020qzy}. It is essential to acknowledge that both angular momentum and magnetic fields, which oppose collapse and consequently extend the actual collapse duration, were disregarded in our calculations. In summary, the principle of Virial equilibrium, denoted by the equation $\langle K\rangle=-\frac{1}{2}\langle U\rangle$, the fundamental concept of the Equipartition theorem, expressed as $\langle K\rangle=\frac{1}{2}k_\text{B}T$, and the influential gravitational potential function $V(r)$, collectively serve as the fundamental pillars underlying the computations of Jeans mass.


Let us now calculate the Jeans mass of a fractional system. It is common knowledge that the fractional extension of a traditional Hamiltonian can be expressed as
\begin{equation} \label{fracclasshamil}
H_\alpha(\textbf{p},\textbf{r})={D_\alpha}|\textbf{p}|^\alpha+V(\textbf{r}), 
\end{equation}
where $D_\alpha$ denotes a generalized coefficient with the
dimension
$[D_\alpha]=erg^{1-\alpha}\,{cm}^\alpha\,{seg}^{-\alpha}$, and
$\alpha$ is the L\'{e}vy's fractional parameter and it is
associated to the concept of L\'{e}vy path \citep{doi:10}. In
addition, the potential term in the above extended Hamiltonian is
given by (\ref{SS15}) in the gravitational case. In the special
case where $\alpha=2$ and $D_\alpha=1/{2m}$, the equation
\eqref{fracclasshamil} reduces to the standard Hamiltonian
$H(\textbf{p},\textbf{r})=\textbf{p}^2/2m+V(\textbf{r})$. In the
realm of gravitational systems, the fractional derivative is
believed to stem from quantum gravity at its most fundamental
level \citep{Rasouli:2022bug, Jalalzadeh:2022uhl,
Jalalzadeh:2021gtq, Rasouli:2021lgy, Moniz:2020emn}. Thus, the
coefficient $D$ is connected to the Planck mass and can be
expressed as
\begin{equation}
    D_\alpha=\frac{(M_\text{P}c)^{2-\alpha}}{2m},
\end{equation}
where $M_\text{P}=\sqrt{\hbar c/G}$ represents the Planck mass.

The equipartition theorem has a generalized version, which can be expressed as \citep{Pathria:1996hda}
\begin{eqnarray}\label{4}
\big<x_m\frac{\partial H}{\partial x_n}\big>=\delta_{mn}k_\text{B}T,
\end{eqnarray}
in which $H$ denotes the system Hamiltonian, a function
of degrees freedom $x_i$. Thus, for an ideal fractional gas whose
Hamiltonian is $H=D_\alpha p^\alpha$, we find
$\big<K\big>=\frac{1}{\alpha}k_\text{B}T$ as the energy per degree of
freedom. Clearly, for $\alpha=2$ and $D_2=1/2m$, the classical ideal gas is recovered. Generally, $H=D_\alpha p^\alpha$ is
accepted as the Hamiltonian of a non-interacting (ideal)
fractional gas for which $\alpha$ represents the fractional
parameter \citep{fractaldimension}. For a general Hamiltonian
$H_=D_\alpha p^\alpha+V(r)=D_\alpha p^\alpha+kr^{l+1}$, where $l$
is a free parameter (as examples, $l=-d$ and $l=-2$ for potential~(\ref{SS15}) and Newtonian potential, respectively), the Virial theorem is also found as
\citep{Laskin:2002zz, virialfractal}
\begin{eqnarray}\label{5}
\langle K\rangle=\frac{l+1}{\alpha}U.
\end{eqnarray}

Now, let us consider a cloud of point masses with constant density, $\rho_0$. Based on Eqs. (\ref{FB5}), (\ref{FB6}), and (\ref{SS10a}), we can express the surface element of a sphere and the volume element in spherical coordinates as
\begin{equation}\label{Surfelement}
    d\mathcal S_\text{fractal}=(4\pi)^\frac{(2-d)}{4}\tilde r^2d\Omega,
\end{equation}
\begin{equation}\label{Volelement}
    d\mathcal V_\text{fractal}=(4\pi)^{\frac{3}{4}(2-d)}\tilde r^2d\tilde rd\Omega=\frac{dL^3}{2}\left(\frac{r}{L}\right)^{\frac{3d}{2}-1}d\left(\frac{r}{L}\right)d\Omega,
\end{equation}
respectively, where $d\Omega=\sin(\theta)d\theta d\phi$ is the line element of a unit 2-sphere.
It is evident that the surface element (\ref{Surfelement}) yields the fractal surface (\ref{FB5}). Also, by utilizing the volume element (\ref{Volelement}), we can determine the mass of the cloud inside a spherical region
\begin{equation}
    \label{Massfrac}
    M_\text{fractal}(r)=M_0\left(\frac{r}{L} \right)^{\frac{3d}{2}},
\end{equation}
where
\begin{equation}
    \label{MassUnit}
 M_0=\frac{4\pi}{3}\rho_0L^3,   \end{equation}
 is the gas mass of the grid volume.


The equation above demonstrates that the mass of the ball region in the fractal space follows a power law relationship, where the fractal dimension of mass in the ball region is equal to
\begin{equation}
    \label{MassD}
    D_\text{mass}=\frac{3d}{2}.
\end{equation}
 Note that \citet{Tarasov:2005orz} introduced a similar power law relation for a fractal mass. {Several years ago, it came to light that the nearby cold interstellar medium possesses a fascinating fractal structure that extends across a vast range of scales, spanning over four orders of magnitude. This remarkable fractal pattern contains a broad spectrum, ranging from immense giant molecular clouds spanning 100 pc to minuscule Bok globules or tiny clumps measuring between 0.01 and 0.1 pc. Furthermore, the tiniest scale of this intricate fractal formation is intricately linked to the critical transition from the isothermal regime to the adiabatic regime. When the temperature plummets to a mere 3 K, the smallest clumps, aptly named clumpuscules \citep{Pfenniger:1996rd}, emerge with a mass of approximately $10^{-3}$ solar masses and an impressive radius of about $L=20$ AU. The discovery of this multifaceted fractal structure within the cold interstellar medium has captivated scientists for years, unveiling the mesmerizing complexity and elegance of the cosmic tapestry surrounding us. The most suitable match for the wide range of observations is when the value of $D_\text{mass}$ is between 1.4 and 2 \citep{deVega:1996gj}. As a result, regarding the relation (\ref{MassD}), we obtain the following restriction on $d$ by observational data
\begin{equation}
    \label{Dex}
    \frac{2.8}{3}\leq d\leq \frac{4}{3}.
\end{equation}
Therefore, gas fragmentation not only leads to a decrease in the value of the fractal dimension $d$ but also expands the grid size from the Planck length in the quantum gravity regime to an estimated range of approximately 20 AU. This implies that the phenomenon of gas fragmentation not only affects the fractal dimension but also significantly impacts the overall scale of the grid, causing it to increase significantly.
}

Under the above circumstances, the cloud of point masses with a constant density  has the following fractal gravitational
potential energy
\begin{eqnarray}\label{10}
U_\text{fractal}=\int_0^R\rho_0V(r)d\mathcal V_\text{fractal},
\end{eqnarray}
where $V(r)$ is the fractional-fractal gravitational potential obtained in (\ref{SS15}).
Integration of the above equation gives us
\begin{equation}
    \label{Energyfract}
    U_\text{fractal}=-\frac{3dG_dM^2}{2(d-1)(2d+1)R^{d-1}},
  \end{equation}
where $G_d$ is defined by Eq. (\ref{SS14}).
Obviously, the Newtonian regime is recovered for $d=2$.

\begin{table}
\centering
\caption{Physical parameters of some Bok globules.The symbol $\textmd{M}_\odot$ denotes the mass of Sun, and the parameters $T$, $M$, and $M^J$ represent the temperature, mass, and Jean's mass of Bok globules \citep{Vainio:2015ejx}, respectively.}
\begin{tabular}{cccc}
\hline
Bok globule & $T$(K) & $\frac{M}{\textmd{M}_\odot}$ & $\frac{M^J}{\textmd{M}_\odot}$  \\ \hline\hline\\
CB $87$ &~$11\cdot4$ &~$2\cdot73$ &~$9\cdot6$ \\
CB $110$ &~$21\cdot8$ &~$7\cdot21$ &~$8\cdot5$ \\
CB $131$ &~$25\cdot1$ &~$7\cdot83$ &~$8\cdot1$ \\
CB $161$ &~$12\cdot5$ &~$2\cdot79$ &~$5\cdot4$ \\
CB $184$ &~$15\cdot5$ &~$4\cdot70$ &~$11\cdot4$ \\
FeSt $1-457$ &~$10\cdot9$ &~$1\cdot12$ &~$1\cdot4$ \\
Lynds $495$ &~$12\cdot6$ &~$2\cdot95$ &~$6\cdot6$  \\
Lynds $498$ &~$11\cdot0$ &~$1\cdot42$ &~$5\cdot7$ \\
Coalsack &~$15$ &~$4\cdot50$ &~$8\cdot1$ \\
\hline\hline
\end{tabular}\label{Tab0}
\end{table}

The Virial theorem (\ref{5}) equips us with
\begin{eqnarray}\label{11}
\langle K\rangle=-\frac{1}{2}(d-1)^2U_\text{fractal},
\end{eqnarray}
Similar to the original model, for the occurrence of collapse to occur, it is essential that the absolute value of the gravitational potential energy (\ref{Energyfract}) surpasses that of the thermal kinetic energy, $K=\frac{3}{2}Nk_\text{B}T$, (where $N=M_\text{fractal}/m$) possessed by the cloud. Consequently, when we substitute the conditions above into the Virial Theorem (\ref{11}), we witness the transformation of the inequality into the equation $Nk_\text{B}T<(d-1)^2U_\text{fractal}/3$. The fractal Jeans  Mass can be found precisely at the threshold of this inequality as
\begin{equation}
   M^\text{fractal}_\text{J}=M_0^\frac{2(1-d)}{2+d}\left(\frac{2(2d+1)(4\pi)^\frac{d-2}{2}Lk_\text{B}T}{d(d-1)mG} \right)^\frac{3d}{d+2}.
\end{equation}

Therefore, each cloud with mass greater than $M_\text{J}^{(\text{fractal})}$ experiences a collapse under the control of fractional gravity (\ref{SS13}), and consequently, a fractional star shall be formed. With a bit of algebra, we have
\begin{equation}
    M^\text{fractal}_\text{J} = M_\text{J}^{\frac{4(d-1)}{2+d}}\left(\frac{5k_\text{B}T}{mG}\right)^{\frac{3(2-d)}{d+2}} L^{\frac{3(2-d)}{2+d}} \left( \frac{2(2d+1) (4\pi)^{\frac{d-2}{2}}}{5d(d-1)} \right)^{\frac{3d}{d+2}}.
\end{equation}
As it is seen, for $d=2$, one can easily find $M_\text{J}^{(\text{fractal})}= M_\text{J}$.

\begin{figure}
    \centering
    \includegraphics[width=8cm]{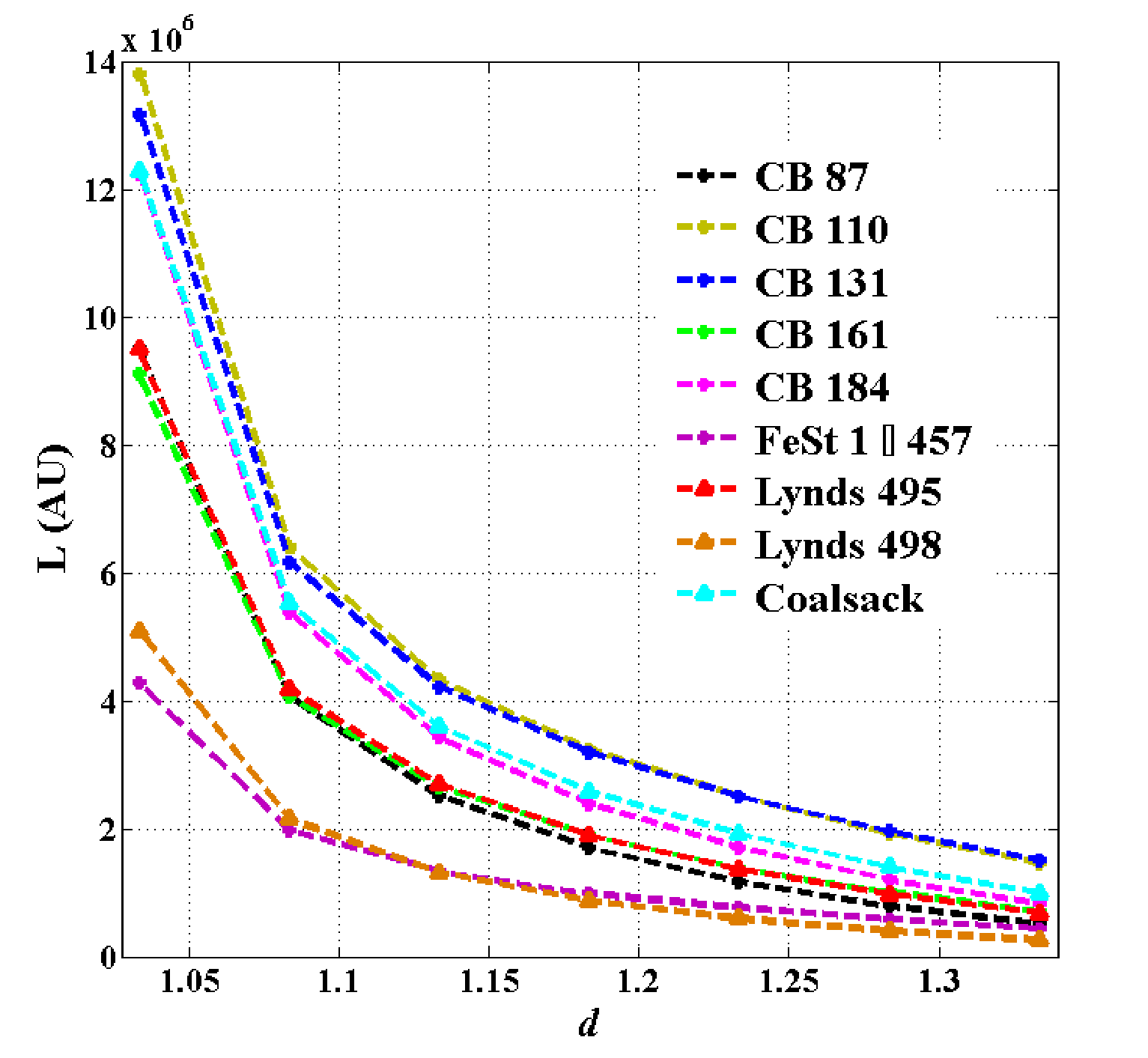}
    \caption{The parameter $L$ versus different values of $d$ obtained for Bok globules given in Table \ref{Tab0}.}
    \label{Fig0}
\end{figure}

\begin{figure}
    \centering
    \includegraphics[width=7.4cm]{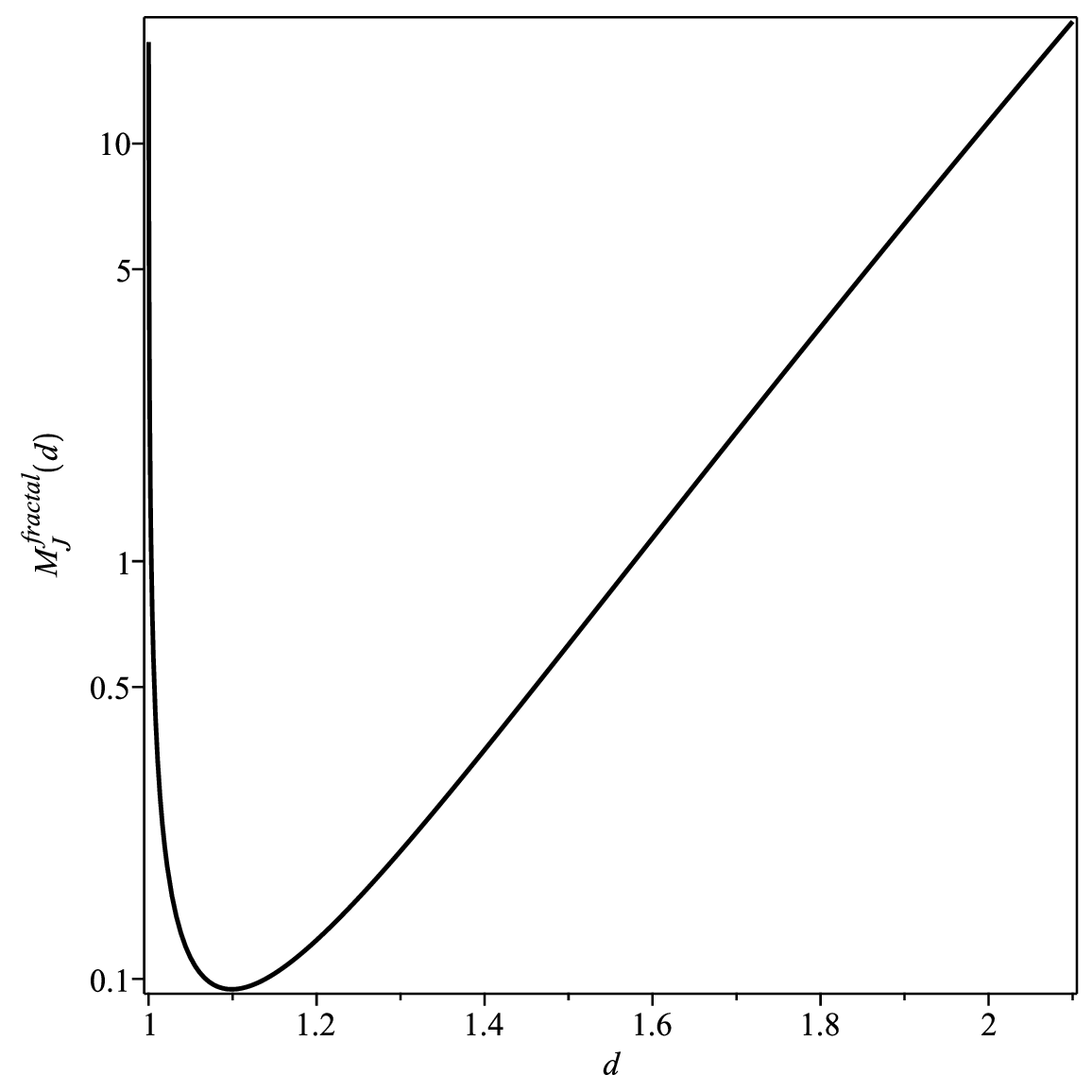}
    \caption{Logplot of the fractal Jeans mass of a cloud as a function of the fractal dimension in units of the sun mass, $M_\odot=1.99\times10^{30}$ kg.}
    \label{Fig1}
\end{figure}

Let us calculate the fractional-fractal Jeans mass of a typical Bok globule. Bok globules, which are characterized as dark clouds consisting of dense cosmic dust and gaseous matter, serve as the breeding grounds for the emergence of new stars. These intriguing formations are primarily located within H II regions, which are regions of ionized gas resulting from the strong ultraviolet radiation emitted by nearby massive stars. With a mass ranging from approximately $2-50M_\odot$, Bok globules encompass a relatively compact region spanning about a light-year in distance. The cloud is mainly made up of hydrogen with number density $n=\rho_0/m=10^4$ atom/cm$^3$ `  mean molecular weight`$\mu=2$, and a temperature of $T=10$ K. {Regarding these conditions, the ordinary Jeans mass (i.e., $d=2$) is $M_J=11.27M_\odot$. If we assume $L=20$ AU, then Eq. (\ref{MassUnit}) gives $M_0=1.9\times10^{-9}M_\odot$.

We listed necessary physical parameters of some Bok globules in Table \ref{Tab0} to obtain $L$. Figure \ref{Fig0} shows the values of $L$ for different possible values of $d$ given in (\ref{Dex}).

As shown in Fig. \ref{Fig1}, the Jeans mass reaches its minimum value $M^\text{fractal}_\text{J}=0.1M_\odot$ for $d=1.1$. This particular value of $d$ is in complete accordance with the range that has been derived from the experimental findings in the inequality (\ref{Dex}). In addition, inserting the value of $d$ into relation (\ref{MassD}) gives us the fractal mass dimension $D_\text{mass}=1.65$ which is very close to its observational value $D_\text{mass}=1.7$ \citep{1981MNRAS809L,Pfenniger:1996rd}}

Our study has revealed that fractal clouds, compared to regular gases with $d=2$, face significantly lower barriers during their collapse process. This suggests that the mass required for a fractal cloud to undergo gravitational collapse is smaller than the ordinary Jeans mass. Essentially, this means that the unique properties of fractal clouds allow them to overcome obstacles more easily, facilitating their collapse.
\section{Burning temperature of fractional stars}

The Gamow theory gives us an estimation of the temperature
of stars \citep{Gamowbook}. Based on this theory, if the mutual
distance of two particles with atomic numbers $Z_1$ and $Z_2$
becomes equal to their de Broglie wavelength, then they can
overcome the Coulomb barrier due to their charge. Indeed, in this
manner, the Kinetic energy of particles equates to the Coulomb
potential as
\begin{eqnarray}\label{13}
\big<K\big>=\big<D_\alpha p^\alpha\big>=\frac{Z_1Z_2e^2}{4\pi\epsilon_0\lambda}
\Rightarrow\lambda=(\frac{4\pi\epsilon_0D_\alpha\hbar^\alpha}{Z_1Z_2e^2})^\frac{1}{\alpha-1},
\end{eqnarray}
in which $\lambda=\frac{\hbar}{p}$ denotes the de
Broglie wavelength. Bearing $\alpha=\frac{2}{d-1}$ in mind, $\lambda$ can also be re-expressed as
\begin{equation}\label{14}
\begin{split}
\lambda=&\lambda_G\lambda_d,\\
\lambda_d=&\bigg[\frac{(2mD_d\hbar)^{\frac{3-d}{d-1}}}{\hbar}
\big(\frac{4\pi\epsilon_0D_d\hbar^2}{Z_1Z_2e^2}\big)^{\frac{2(d-2)}{d-1}}\bigg]^\frac{d-1}{3-d},
\end{split}
\end{equation}
that produces the Gamow original calculations
for $d=2$, where $D_2=\frac{1}{2m}$ \citep{Gamowbook,
MoradpourJava}. Here,
$\lambda_G=\frac{2\pi\epsilon_0\hbar^2}{mZ_1Z_2e^2}$ represents the
de Broglie wavelength whenever $d=2$ ($\lambda_2=1$). On the other hand, in a fractional space of $d$ fractal dimension, we have $3\frac{d}{2}$ degrees of freedom combined with Eq.~(\ref{4}) and the above
equations to finally reach
\begin{eqnarray}\label{15}
T_d=\frac{2}{\lambda_dd(d-1)}T_G,
\end{eqnarray}
where $T_G=\frac{Z_1Z_2e^2}{6\pi\epsilon_0k_\text{B}\lambda_G}$
is the ordinary Gamow temperature ($d=2$) \citep{Gamowbook,
MoradpourJava}.

\section{{Hydrostatic Equilibrium}}\label{sec:Methods}

In the context of fractional calculus, Eq. (\ref{SS13}) describes the gravitational force ($F_g$) between two masses ($m$ and $M$). To understand the motion of objects under gravitational influence, let us consider a star with a total mass ($M$) and a radius ($r$). Specifically, we will concentrate on a thin layer of matter located at a distance of $r$ from the star's core. This layer has a thickness of $dr$ and contains gas with a density of $\rho$. Consequently, the shell's mass can be determined by
\begin{eqnarray}\label{Conti}
dm = \rho~d\mathcal V_\text{fractal}.
\end{eqnarray}
The shell experiences two different types of forces. The initial force acting on it is due to gravity $F_g$. If we define $m$ as the star's mass within radius $r$, we can observe that  $\rho dr$ represents the mass of the shell per cross-sectional area. According to the fractional gravitational force~(\ref{SS13}), the force of gravity per unit area exerted on the shell can be described as follows
\begin{eqnarray}\label{Fgshell}
F_g = - G_{d} \frac{m}{r^d}\rho dr.
\end{eqnarray}
The presence of a minus sign indicates that the force per area is directed inward.

The shell also experiences a force per area due to gas pressure. It is important to note that the shell only feels a net pressure force when there is a difference in pressure on either side. If the pressure at the base of the shell is denoted as $P(r)$ and the pressure at the top is $P(r + dr)$, the shell experiences a net pressure force per area as follows
\begin{eqnarray}\label{Fpshell}
F_p = P(r) - P(r + dr).
\end{eqnarray}
It is important to note that the sign convention used here ensures that the force exerted by the top of the shell, represented by the term $P(r + dr)$, is directed inward, while the force from the bottom of the shell is directed outward. As the limit of $dr$ approaches zero, it becomes advantageous to express this term more clearly by utilizing the definition of the derivative:
\begin{eqnarray}\label{HE01}
\frac{dP}{dr} = \lim_{dr \rightarrow 0} \frac{P(r + dr) - P(r)}{dr}.
\end{eqnarray}
Using this, the pressure force per area is obtained as
\begin{eqnarray}\label{HE02}
F_p = -\frac{dP}{dr} dr.
\end{eqnarray}
Based on Newton's second law, the net force per area experienced
by the shell equals $F = dm~a$, wherein $a$ is acceleration, and
$dm = \rho~dr$, and hence, we can write the following relation for
the shell
\begin{eqnarray}\label{HE03}
(\rho~dr)a = - \left(G_{d} \frac{m}{r^d}\rho + \frac{dP}{dr}\right) dr.
\end{eqnarray}
The following equation provides information on the shell's acceleration in relation to the forces acting upon it. It represents the equation of motion for the shell.
\begin{eqnarray}\label{HE04}
a = - G_{d} \frac{m}{r^d} - \frac{1}{\rho} \frac{dP}{dr}.
\end{eqnarray}
The condition of hydrostatic equilibrium requires that $a = 0$. Therefore, we obtain
\begin{eqnarray}\label{HE05}
G_{d} \frac{m}{r^d}\rho = - \frac{dP}{dr}.
\end{eqnarray}
This equation is commonly referred to as the hydrostatic
equilibrium equation, as it signifies the state where the star
maintains the static pressure balance.

Can we predict how fast a shell of material within a star will
accelerate? To answer this question, we can consider a similar
situation with a significant difference between the pressure per
unit area and gravitational forces, leading to substantial
acceleration. If the pressure were negligible, the outermost shell
would experience free-fall motion towards the star's center due to
gravity. In this case, we can quickly determine how long it would
take for the star to change its size or other properties. The
velocity at which it would accelerate is determined by the
characteristic free-fall speed $v_{ff}$ provided by gravity as
follows
\begin{eqnarray}\label{ffs}
v_{ff} = \sqrt{\frac{2~M~G_d}{(d-1)R^{d-1}}},
\end{eqnarray}
where $M$ and $R$ are the mass and radius of the star,
respectively. Moreover, the time it would require for the shell to
descend toward the star's center is determined by dividing the
distance to the center by its velocity. This time interval is
known as the star's dynamical $t_{dyn}$ timescale, representing
the duration it would take for the star to reorganize without a
balance between pressure and gravity. Therefore, for the timescale
$t_{dyn}$ (or equally, the time required for star formation), we
have
\begin{equation}\label{tdyn}
t_{dyn}= \frac{R}{v_{ff}} \approx \sqrt{\frac{R^{1-\frac{d}{2}}}{\bar{\rho}G_d}},
\end{equation}
where $\bar{\rho}=M/\mathcal V_\text{fractal}$ is the mean density of a typical star with radius $R$.

Using hydrostatic equilibrium (i.e., Eq.~(\ref{HE05})), we can write
\begin{eqnarray}\label{LM01}
 \frac{d}{dr}\left(\frac{1}{\rho}\frac{dP}{dr} \right) =-\frac{d}{dr}\left(G_{d} \frac{m}{r^d} \right)= G_{d} \frac{md}{r^{d+1}} - \frac{G_d}{r^d} \frac{dm}{dr},
\end{eqnarray}
combined with Eq.~(\ref{Conti}) to finally obtain
\begin{align}\label{LM03}
\frac{d}{dr}\left(\frac{1}{\rho}\frac{dP}{dr} \right) +
\frac{d}{r\rho} \frac{dP}{dr} +2d\pi G_d\left(\frac{r}{L^3}
\right)^{\frac{d}{2}-1}\rho=0.
\end{align}
\noindent Here, we used Eq.~(\ref{HE05}) to write the
term $G_{d} \frac{md}{r^{d+1}}$ as $-\frac{d}{r\rho}
\frac{dP}{dr}$. Eq.~(\ref{Conti}) has also been employed to
rewrite $\frac{G_d}{r^d}\frac{dm}{dr}$ as $\frac{G_d}{r^d}\rho
\frac{d\mathcal V_\text{fractal}}{dr}$ leading to
$2d\pi\frac{G_d}{r^d}\rho\frac{r^{3\frac{d}{2}-1}}{L^{3(\frac{d}{2}-1)}}$ that finally generates Eq.~(\ref{LM03}). By
taking $d=2$, Eq.~(\ref{LM03}) is converted to the well-known
Lane--Emden equation obtained using Newtonian force. Concerning
$\rho = \rho_{c}\Theta^n$ and the polytropic equation of state $P
= K \rho^{1+\frac{1}{n}}$, we have $P =
K\rho_{c}^{1+\frac{1}{n}}\Theta^{n+1}$, wherein $K$ is a constant,
$\rho_{c}$ is the central density of a star, $n$ is the polytropic
index, and $\Theta$ is a function related to density. Thus, the
following differential equation is finally reached
\begin{align}\label{LMN02}
\frac{1}{\xi^{\frac{3d}{2}-1}} \frac{d}{d\xi} \left(\xi^d \frac{d\Theta}{d\xi}\right) + \Theta^n = 0,
\end{align}
 where
\begin{align}\label{LMN03}
\xi\equiv\frac{r}{\psi},~~~~~~ \psi\equiv\left(\frac{K(n+1)\rho_c^{\frac{1}{n}-1}}{2\pi d G_d L^{3(1-\frac{d}{2})}}\right)^{\frac{1}{1+\frac{d}{2}}},
\end{align}
and it is easy to check that $d=2$ leads to the results obtained in the Newtonian regime. It seems that only $n=0$ and $n=1$ have analytical solutions as
\begin{align}\label{S0}
\Theta(\xi)=c_1 \frac{\xi^{1-d}}{1-d} + c_2 - \frac{4\xi^{\frac{d}{2}+1}}{3d^2+6d},
\end{align}
and
\begin{multline}\label{S1}
\Theta(\xi)=c_3~ 2^{\frac{2(1-d)}{d+2}} (d+2)^{\frac{2(d-1)}{d+2}} \xi^{\frac{1-d}{2}}~ \Gamma\left(\frac{3d}{d+2}\right) J_{\frac{2(d-1)}{d+2}} \left(\frac{4 \xi^{\frac{d+2}{4}}}{d+2}\right) + \\
c_4~ (\frac{d}{2}+1)^{\frac{2(d-1)}{d+2}} \xi^{\frac{1-d}{2}}~ \Gamma\left(\frac{4-d}{d+2}\right) J_{-\frac{2(d-1)}{d+2}} \left(\frac{4 \xi^{\frac{d+2}{4}}}{d+2}\right)
\end{multline}
respectively. In these equations, $c_i$ ($i=1,\cdots,4$) are
constants of integration, $\Gamma(d)$ is the gamma function,
$J_k(\xi)$ denotes the Bessel functions of the first kind, and
again, the results of the Newtonian regime are recovered by
inserting $d=2$.

Therefore, we have successfully derived analytical solutions for a
specific class of polytropic models by solving the corresponding
Lane--Emden equation for fractional stars. Our models are
characterized by a power-law equation of state that definitively
establishes the relationship between pressure and density.

\begin{figure*}
\centerline{\includegraphics[width=1\textwidth,clip=]{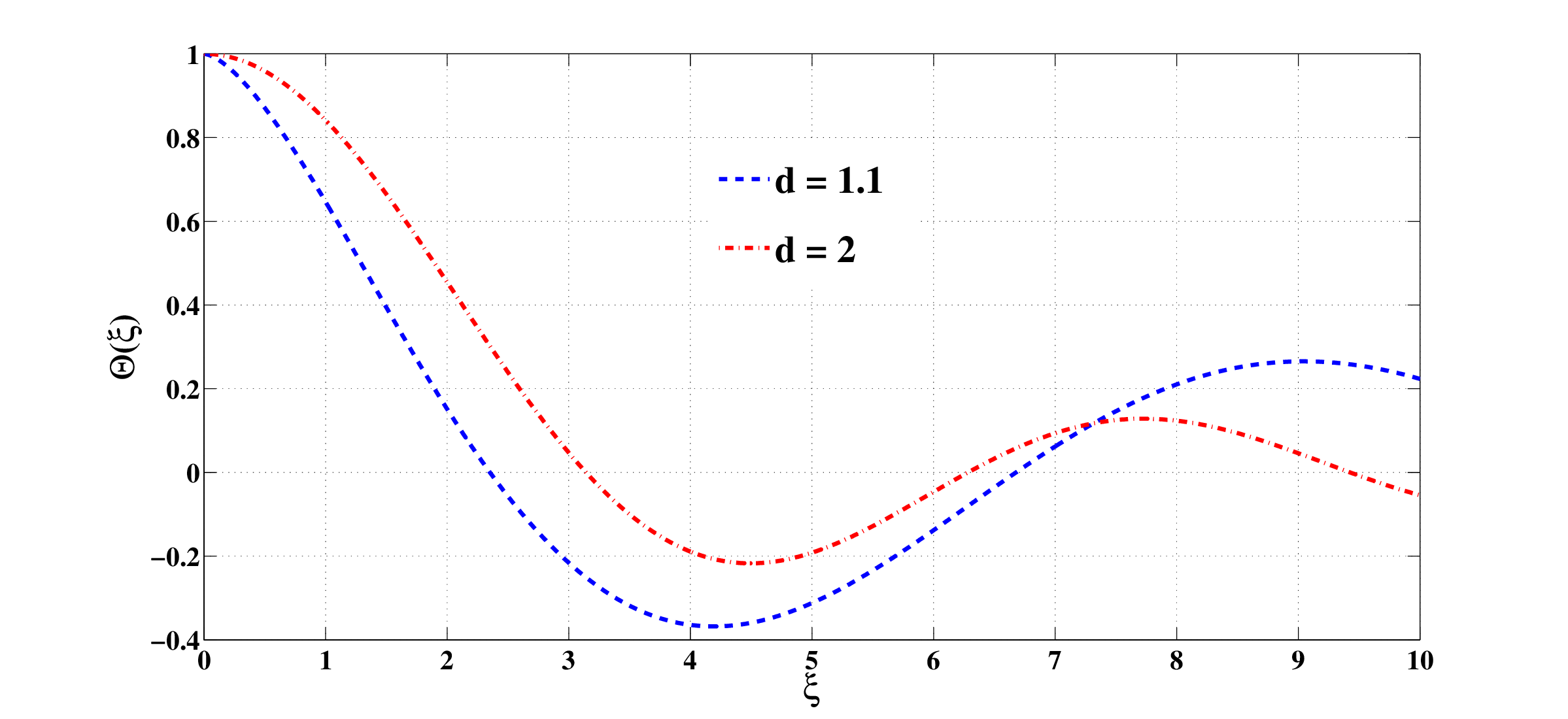}}
\caption{Dimensionless parameter $\Theta$ versus $\xi$ for $n=1$, and two values of $d=1.1$ and the Newtonian limit $d=2$.}\label{SizeallHigh}
\end{figure*}

In the fractional Lane--Emden equation, we can rewrite it as a
system of first-order ordinary differential equations (ODEs) by
introducing a new variable $z = \frac{d\Theta}{d\xi}$. The system
becomes:
\begin{eqnarray}\label{FracLaneEmden}
z = \frac{d\Theta}{d\xi},~~~~~~~~~~~~~~\nonumber\\
\frac{dz}{d\xi} = - \frac{\Theta^n}{\xi^{1-\frac{d}{2}}} - \frac{zd}{\xi}.
\end{eqnarray}
This system is then used in the "ode45" solver to numerically
integrate the fractional Lane--Emden equations over a specified
range of $\xi$. The value "1" is considered for the initial value
of $\Theta$ at the starting point $\xi = \xi_0$. In many problems,
$\Theta$ is often normalized to "1" at the centre, which is a
common initial condition. The value "0" is considered for the
initial value of $\frac{d\Theta}{d\xi}$ (or $z$) at the starting
point $\xi = \xi_0$. This is often set to "0", assuming a regular
solution at the centre. Together, these initial conditions express
that at the starting point $\xi = \xi_0$, the function $\Theta$
has a value of "1", and its derivative $\frac{d\Theta}{d\xi}$ is
"0". As an example, the numerical solution of Eq. (\ref{LMN02})
is plotted ($\Theta$ versus $\xi$) in Fig. \ref{SizeallHigh} for $d=1.1$ and the Newtonian regime ($d=2$) to have a comparison.

The star mass is calculate-able using the integral
\begin{eqnarray}\label{57}
M=\int_0^R\rho_0d\mathcal V_\text{fractal},
\end{eqnarray}
\noindent combined with the lane--Emden equation to finally reach at
\begin{eqnarray}\label{58}
M=2d\pi\frac{\rho_c\psi^{3\frac{d}{2}}}{L^{3\frac{d}{2}-3}}\bigg[-\xi^d\frac{d\Theta}{d\xi}\bigg]_{\xi=0}^{\xi=\frac{R}{\psi}}.
\end{eqnarray}
Now, bearing in mind the boundary conditions $\Theta=1$ and $\frac{d\Theta}{d\xi}=0$ at $\xi=0$, one finally obtains
\begin{eqnarray}\label{59}
M=2d\pi\frac{\rho_c\psi^{3\frac{d}{2}}}{L^{3\frac{d}{2}-3}}\bigg[-\xi^d\frac{d\Theta}{d\xi}\bigg]_{\xi=\frac{R}{\psi}},
\end{eqnarray}
recovering the Newtonian regime when $d=2$. Therefore, the value of $\frac{d\Theta}{d\xi}$ at $\xi=\frac{R}{\psi}$ is important to estimate the mass and in parallel, for each $n$, the maximum value of $M$ also depends on the values of $d$ and $L$. As en example, bearing in mind the mentioned boundary conditions, Eq.~(\ref{LMN02}) leads to $-\xi^d\frac{d\Theta}{d\xi}=\frac{2\xi^{3d/2}}{3d}$ for $n=0$, and thus
\begin{eqnarray}\label{60}
M=\frac{4\pi L^3}{3}~\rho_c~(\frac{R}{L})^{3d/2},
\end{eqnarray}
that produces the common solution ($M=\frac{4\pi R^3}{3}\rho_c$) when $d=2$ (the Newtonian regime).

By introducing a non-integer free parameter, denoted as $d$, based
on fractional calculus, we can obtain additional solutions for the
structure of stars beyond the integer-based ones. This is
illustrated in Fig. \ref{Fig1}. By incorporating this non-integer
parameter into the new Lane--Emden equation, we can obtain a final
solution for the density profile as a function of radius. We
derived our equations through standard algebra based on previous
research. These solutions offer a wider range of possibilities for
describing stellar structures, eliminating the need for new
theories or adjustments to stellar structure equations without
substantial cause, such as altering the equation of state. Further
studies require verifying the compatibility of the results with
observations, which can be improved by considering other
Lane--Emden based equations, such as the Emden--Chandrasekhar
equation and Chandrasekhar's white dwarf equation. These studies
are outside the scope of the current work.

Focusing on solutions, we were able to express the solutions using
special functions, such as the polytropic function and polytropic
index. Utilizing special functions, such as the polytropic
function, allowed us to describe the physical properties of the
polytropic models concisely and mathematically. These functions
provide a way to represent the behaviour of the corresponding
equation for the specific class of polytropic models under
consideration. By incorporating the polytropic index, we were able
to capture the dependence of the solutions on the specific
physical parameters of the polytropic models, providing a more
comprehensive understanding of their behaviour. It is important to
note that our analytical solutions were limited to a specific
range of the polytropic index, $n$. In order to explore solutions
for higher orders of $n$ and investigate other classes of models,
we acknowledge the need for numerical methods and further research
efforts. By combining analytical and numerical approaches, future
studies can delve deeper into the properties and behaviours of
polytropic models beyond the scope of our specific class. This
would contribute to a more comprehensive understanding of
astrophysical systems and their dynamics and pave the way for
potential applications in various fields such as stellar
structure, galactic dynamics, and cosmology.

\section{Conclusion}

Astrophysics has shown that there are consistent scaling relations
for mass fluctuation and internal velocity dispersion in different
types of molecular clouds. These scaling relations have been
observed in numerous molecular clouds, indicating a uniform
hierarchical structure, regardless of the scale at which these
clouds are observed, ranging from $10^{-2}$ to 100 pc. However,
despite the extensive empirical evidence supporting these scaling
relations, a complete theoretical understanding of the physical
mechanisms responsible for their emergence and precise
interpretation still needs to be revised. Therefore, there have
been various proposed explanations and contentious discussions
within the scientific community regarding the theoretical
derivation of these scaling relations.

The physics behind the scaling relations observed within molecular
clouds is usually explained as follows. Any amount of gas that is
gravitating and in a state of quasi-isothermal equilibrium within
a thermal bath, assuming that it does not possess excessive
kinetic energy to expand infinitely, tends to adhere to an
isothermal distribution. This means that the gas distribution
tends to be the same temperature throughout. In other words, it
manifests as a sphere where the density diminishes as the distance
from the center increases. On the contrary, the gravothermal
catastrophe occurs when the disparity in density between the
central region and the margin surpasses the critical value of 32.1
\citep{1968MNRAS.95L}. At this particular juncture, the sphere
becomes unstable and breaks into ten distinct clumps while the
outer regions evaporate, thereby acquiring energy from the
surrounding background thermal bath. This inherent instability
repeatedly manifests at the subsequent smaller scale, persisting
as long as the gas maintains its isothermal state, thereby giving
rise to a hierarchical self-similar fractal configuration that is
spontaneously established. The iterative process culminates when
the scale at which the isothermal nature of the system
disintegrates is reached, thereby terminating the recursion. Our
findings suggest that the fractal pattern of ISM clouds may be
primarily driven by gravity, along with gravothermal catastrophe.
This implies that the origin of these structures is more
fundamental than previously assumed and may have significant
implications for our knowledge of the physics of the ISM.

In this article, we have utilized the concept of fractional
gravity, considered a valuable theoretical tool, to explain the
complex nature of scaling relations and hierarchical self-similar
fractal configurations commonly observed within molecular clouds.
This enhances our understanding and knowledge in this fascinating
field of study. The thermodynamics, statistics, and symmetry of
fractional molecular clouds differ from those of ordinary clouds
supported by Newtonian gravity. Indeed, as our calculations
confirm, the gravitational force in a fractional sample is
different from the well-known Newtonian force. Therefore, when the
primary cloud carries a fractional hierarchy, star formation and
star life should be studied using fractional measures. In this
line, the Jeans mass of a fractional cloud was calculated,
indicating that a fractional molecular cloud whose mass is
significantly smaller than $M^J$ can also collapse. In addition,
our analysis further demonstrates that the fractal mass dimension,
$D_\text{mass}=1.65$, exhibited by the model under investigation
aligns with the observations.

By delving into hydrostatic equilibrium and performing
calculations to determine the Gamow temperature, we developed the
Lane--Emden equation for fractional stars. Our perspective on
deriving these equations not only broadens our understanding of
how stars behave in fractional form but also presents a new method
for exploring the dynamics and characteristics of astrophysical
systems. Exploring fractional stars can yield valuable insights
into the physics and phenomena that drive them and has the
potential to make reasonable contributions to the field of
astrophysics, shedding new light on fractional stars and their
place in the broader context of the cosmos. This research has
revealed new details about stars that form within clouds with
masses smaller than the conventional Jeans mass, such as those
found in Bok globules. Our enhanced understanding of fractal
clouds and their distinctive characteristics has allowed us to
explain the collapse process and their crucial role in forming
stars.

While we have gained some insights, further research and careful
observations are necessary to fully comprehend the fractional
aspects of these types of clouds. This paper has presented readers
with an introduction to these stars, establishing the basics for
future exploration.






\bmhead{Acknowledgements}
 S.J. acknowledges financial support from the National Council for Scientific and Technological Development, Brazil--CNPq, Grant no. 308131/2022-3.

\bibliography{Paper}

\end{document}